\documentstyle[epsfig]{article}

\newcommand{\be}{\begin{equation}}
\newcommand{\ee}{\end{equation}}
\newcommand{\beq}{\begin{eqnarray}}
\newcommand{\eeq}{\end{eqnarray}}
\newcommand{\bear}{\begin{array}}
\newcommand{\ear}{\end{array}}

\begin{document}
\setlength{\unitlength}{1mm} {\hfill DSF-12/2004}

\hfill hep-th/0407257\\
\begin{center}
{\Large\bf Space-time symmetry restoration in cosmological models with
Kalb--Ramond and scalar fields}\\
\end{center}

\bigskip\bigskip

\begin{center}
{{\bf E. Di Grezia}, {\bf G. Mangano} and {\bf G. Miele}}
\end{center}
\begin{center}
{\it Dipartimento di Scienze Fisiche, Universit\`{a} di Napoli ``Federico II'',
and INFN, Sezione di Napoli, Complesso Universitario di Monte S. Angelo,
Via Cinthia, I-80126 Napoli, Italy}
\end{center}

\begin{abstract}
We study symmetry of space-time in presence of a minimally coupled scalar
field interacting with a Kalb--Ramond tensor fields in a homogeneous but
initially anisotropic universe. The analysis is performed for the two
relevant cases of a pure cosmological constant and a minimal quadratic,
renormalizable, interaction term. In both cases, due to expansion, a
complete spatial symmetry restoration is dynamically obtained.
\end{abstract}
\bigskip

The most recent measurements of the anisotropy of cosmic background
radiation (CBR) \cite{Bennett:2003bz} strongly support the inflationary
scenario for the dynamics of the early universe, where a scalar field,
dominating the energy density of the primordial plasma, drives the fast
expansion of the universe scale factor. If inflation occurs at very high
energy, it is likely that this phenomenon experiences the non trivial
structure of space time, if any, as the one predicted by noncommutative
geometry. A simple way to implement effectively such a noncommutative
structure has been proposed in Ref.s \cite{Lizzi:2002ib, DiGrezia:2003ug},
where an antisymmetric tensor (Kalb--Ramond) field takes into account the
non-vanishing commutator of space-time coordinates.

In the present letter, we analyze the behaviour of a homogeneous
but initially anisotropic universe, filled with a Kalb--Ramond
tensor $\theta_{\mu\nu}$ which interacts with a scalar field,
minimally coupled with gravity. In particular we look for
space-time symmetry restoration, as an effect of the matter fields
evolution, for two relevant cases, namely a pure cosmological
constant and a quadratic renormalizable interaction term. For such
an involved physical system, the knowledge of the constants of
motion is particularly useful in order to solve the dynamics and
in this concern, the procedure described in Ref.s
\cite{deRitis:ba, deRitis:tp, Capozziello:bi, dem} to find the
Noether symmetries can be applied.

In order to describe a field dynamics in a homogeneous but
spatialy anisotropic space-time, the appropriate metric is a
Bianchi I. In this case the line element can be written as
\begin{equation}
d s^2 = dt^2 - \sum_{i=1}^{3} a_i^2(t) \, (dx^{i})^2 \,\,\, ,
\label{1}
\end{equation}
where $a_i$ is the scale factor for the $i$-th spatial direction. The
corresponding nonvanishing connection coefficients are (no summation over
$i$ is here meant)
\begin{equation}
\Gamma_{ij}^0= \delta_{ij} \, \dot a_i a_i \,\,\, ,~~~~~~
\Gamma_{~0i}^i = \frac{\dot a_i}{a_i}\,\,\, ,
\label{2}
\end{equation}
and for the Ricci tensor one has
\begin{equation}
R_{0}^0= -\sum_{i=1}^{3} \frac{\ddot a_i}{a_i}\,\,\, ,~~~~~~R_{i}^j= -
\delta_i^j \left( \frac{\ddot a_i }{a_i}+\frac{\dot a_i}{a_i}
\sum_{k\neq i} \frac{\dot a_k}{a_k} \right)\,\,\, .
\label{3}
\end{equation}
We consider a model with an antisymmetric tensor, $\theta_{\mu \nu
}$, responsible for the noncommutativity of space-time, and a
minimally coupled scalar field, $\varphi$, which drives the
inflation \cite{DiGrezia:2003ug}. The corresponding action reads
\begin{equation}
S \equiv \int  d^{4} x \sqrt{-g} \textrm{ }\left[-\frac{R}{16\pi
G}+\frac{1}{12} \textrm{H}_{\mu \nu \sigma}\textrm{H} ^{\mu \nu
\sigma} +\frac{1}{2} \varphi _{;\mu }\varphi ^{;\mu }-V(\varphi
)- W(\varphi, \theta_{\mu \nu}\theta^{\mu \nu})\, \right],
\label{action}
\end{equation}
where
\begin{equation}
H_{\mu \nu \sigma}  \equiv \nabla _{\mu }\theta _{\nu \sigma }+
\nabla _{\nu }\theta _{\sigma \mu } +\nabla_{\sigma }\theta _{\mu
\nu }\,\,\,\, , \label{1a}
\end{equation}
is the field strength associated to the antisymmetric tensor
$\theta _{\mu \nu }= -\theta _{\nu \mu }$.\\ By using the
space-time homogeneity we can safely assume that all fields depend
on time only. Moreover, one can show that $\theta ^{0i}=\theta
_{0i}=0$ and that it is possible to take only one component of
$\theta_{ij}$ to be non vanishing, like for example
$\theta_{12}=\theta$ \cite{DiGrezia:2003ug}. Indeed this condition
guarantees that the energy momentum tensor is diagonal. Therefore
it should be imposed as a constraint since we assume a Bianchi I
metric. In this case the Lagrangian density results
\begin{eqnarray}
{\cal L} &=& - \, M^2 \left( a_{1}\dot{a}_{2}\dot{a}_{L} +
a_{2}\dot{a}_{1}\dot{a}_{L} + a_{L}\dot{a}_{1}\dot{a}_{2} \right)
\nonumber\\ &+& a_{1}a_2 a_{L}\left[ \frac{\dot{\varphi}^2}{2} - V(\varphi)-
W\left(\varphi,\frac{\theta^2}{a_{1}^{2}a_{2}^{2}}\right) + \frac {1}{2}
\,\frac{\dot{\theta}^2}{a_{1}^{2}a_{2}^{2}} \right]\,\,\, , \label{lan1}
\end{eqnarray}
where according to our notations $a_L \equiv a_3$ and $M^2=(8\pi
G)^{-1}$.\\ As in \cite{deRitis:ba, deRitis:tp,Capozziello:bi,dem}, we
investigate the Noether symmetries of ${\cal L}$ which can help in solving
the dynamical system. We restrict our analysis to those transformations on
the tangent space which preserve the second order character of field
dynamics. To this aim, let us consider a generic vector field $\bf{X}$ on
the tangent space TQ
\begin{equation}
 \bf{X}=\alpha_1 \frac{\partial}{\partial
a_{1}}+ \alpha_2 \frac{\partial}{\partial a_{2}} + \beta \frac
{\partial}{\partial a_{L}} + \gamma \frac {\partial}{\partial
\varphi} +\delta \frac {\partial}{\partial \theta}+
\dot{\alpha}_1\frac {\partial}{\partial \dot{a}_{1}} +
\dot{\alpha}_2\frac {\partial}{\partial \dot{a}_{2}}+
\dot{\beta}\frac {\partial}{\partial \dot{a}_{L}}+
\dot{\gamma}\frac{\partial}{\partial \dot{\varphi}}+
\dot{\delta}\frac {\partial}{\partial \dot{\theta}}
\end{equation}
and such that the corresponding Lie derivative vanishes
\begin{equation}
L_X {\cal L}= 0 \, \, \, . \label{lie2}
\end{equation}
It is easy to see that condition (\ref{lie2}) leads to the following set of
differential equations
\begin{eqnarray}
&&\left(\frac{\alpha_1}{a_1} +  \frac{\alpha_2}{a_2} +
\frac{\beta}{a_{L}} \right)(V + W) + \gamma (V_{,\varphi} +  W_{,\varphi})
+ \delta \, W_{,\theta}+ \alpha_1 \, W_{,a_1} +
\alpha_2 \, W_{,a_2}
=0\,\,\, ,\nonumber\\\label{ers}\\
&& \alpha_1 \, a_{2}\,a_{L}+ \alpha_2 \,a_{1}\,a_{L} + 2\,
 a_1\,a_2\,a_{L}
\frac {\partial\gamma}{\partial \varphi} + \beta \, a_1 \, a_2=0 \,\,\, ,
\label{er1s}\\
&&-\alpha_1 \,a_{2}\,a_{L}- \alpha_2 \,a_{1}\,a_{L}+ \beta \,a_1\,
a_2 + 2\,a_{L}\,a_1\,a_2 \frac {\partial\delta}{\partial
\theta}=0 \,\,\, ,\label{er2s}\\
 &&a_3 \, \frac{\partial\alpha_2}{\partial a_1} + a_2 \, \frac {\partial\beta}
 {\partial a_{1}} =0\,\,\, ,\label{er3s}\\
 && a_{3} \, \frac
{\partial\alpha_1}{\partial a_{2}} + a_{1} \, \frac
{\partial\beta}{\partial a_{2}}=0 \,\,\, , \label{er4s}\\
&& a_{2} \, \frac {\partial\alpha_1}{\partial a_{3}} + a_{1} \,
\frac
{\partial\alpha_2}{\partial a_{3}}=0 \,\,\, , \label{er4biss}\\
&& a_{L} \, \frac {\partial\alpha_2}{\partial \varphi}+ a_{2}\,
\frac {\partial\beta}{\partial \varphi}-\frac{a_{1} \, a_2 \,
a_{L}}{M^2} \, \frac{\partial\gamma}{\partial a_{1}}=0\,\,\,
,\label{er5s}\\
&& a_{L} \, \frac {\partial\alpha_1}{\partial \varphi}+ a_{1} \,
\frac {\partial\beta}{\partial \varphi} - \frac{a_{1} \, a_2 \,
a_{L}}{M^2} \, \frac {\partial\gamma}{\partial a_{2}}=0\,\,\,
,\label{er5biss}\\
&& a_2 \, \frac {\partial\alpha_1}{\partial \varphi} + a_1 \,
\frac {\partial\alpha_2}{\partial \varphi} - \frac{a_{1} \, a_2 \,
a_{L} }{M^2} \frac {\partial\gamma}{\partial a_{L}} = 0\,\,\,
,\label{er6s}\\ && a_{L} \, \frac {\partial\alpha_2}{\partial
\theta}+ a_{2} \, \frac {\partial\beta}{\partial \theta} -
\frac{a_{L}}{a_1 \, a_{2} \, M^2} \, \frac
{\partial\delta}{\partial a_{1}}=0\,\,\,
,\label{er7s}\\
&& a_{L} \, \frac {\partial\alpha_1}{\partial \theta}+ a_{1} \,
\frac {\partial\beta}{\partial \theta} - \frac{a_{L}}{a_1 \, a_{2}
\, M^2} \, \frac {\partial\delta}{\partial a_{2}}=0\,\,\,
,\label{er7biss}\\
&& a_{2} \, \frac {\partial\alpha_1}{\partial \theta}+a_{1} \,
\frac {\partial\alpha_2}{\partial \theta}  - \frac{a_{L}}{a_1 \,
a_{2} \, M^2} \,  \frac {\partial\delta}{\partial a_{L}}=0\,\,\,
,\label{er8s}\\
&&\frac {1}{a_1 \, a_2} \, \frac {\partial\delta}{\partial
\varphi} +
a_1\, a_2 \, \frac {\partial\gamma}{\partial \theta}=0\,\,\, ,\label{er9s}\\
&&\alpha_1 + a_{2} \, \frac {\partial\alpha_1}{\partial a_{2}} +
a_{L} \, \frac {\partial\alpha_1}{\partial a_{L}} +  a_{1} \,
\frac {\partial\alpha_2}{\partial a_{2}}+ a_{1} \, \frac
{\partial\beta}{\partial a_{3}}=0\,\,\, , \label{tres}\\
&&\alpha_2 + a_{2} \, \frac {\partial\alpha_1}{\partial a_{1}} +
a_{1} \, \frac {\partial\alpha_2}{\partial a_{1}}+ a_{L} \, \frac
{\partial\alpha_2}{\partial a_{L}} +  a_{2} \, \frac
{\partial\beta}{\partial a_{L}}=0\,\,\, , \label{trebiss}
\\&&\beta + a_{3} \frac {\partial\alpha_1}{\partial a_{1}} + a_{3}
\frac {\partial\alpha_2}{\partial a_{2}}+ a_{1} \frac
{\partial\beta}{\partial a_{1}} +  a_{2} \frac
{\partial\beta}{\partial a_{2}}=0\,\,\, . \label{tretriss}
\end{eqnarray}
It is possible to show that the general solution of Eqs.
(\ref{ers})-(\ref{tretriss}) is given by
\begin{eqnarray}
\alpha_1 &=& A \, a_{1}\,\,\, ,\\ \alpha_2 &=& -A \, a_{2}\,\,\, ,\\\beta &=& 0 \,\,\, ,\\
\gamma&=&B \,\,\, ,
\\
\delta&=&C\,\,\, ,
\end{eqnarray}
where $A,B,C$ are arbitrary constants. Thus the general vector fields
$\bf{X}$, which satisfies $L_X{\cal L}=0$ is given by
\begin{eqnarray}
{\bf X}= A \, {\bf X_{\perp}} + B \, {\bf X_\varphi} +C \, {\bf X_{\theta}}
\,\,\, ,\label{x1}
\end{eqnarray}
where the independent and commuting vectors are
\begin{eqnarray}
{\bf X_{\perp}} = a_{1}\frac{\partial}{\partial a_{1}}-  a_{2}
\frac{\partial}{\partial a_{2}} +
\dot{a}_{1}\frac{\partial}{\partial \dot{a}_{1}} -
\dot{a}_{2}\frac{\partial}{\partial \dot{a}_{2}}\,,\,\,\,\,\,\,\,\,\,
 {\bf X_\varphi} = \frac{\partial}{\partial \varphi}\,,\,\,\,\,\,\,\,\,\,
 {\bf X_{\theta}} = \frac{\partial}{\partial \theta}\,.
\end{eqnarray}
with the constraint
\begin{equation}
A \,( a_1 \, \Omega_{,\, a_1} - a_2 \, \Omega_{, \, a_2}) + B \,
\Omega_{,\, \varphi} + C \, \Omega_{,\, \theta} =0\,\,\, ,
\label{comp0}
\end{equation}
where $\Omega = V + W$. Note that from covariance
$\Omega=\Omega\left(\varphi,\theta_{12}\theta^{12}\right)=
\Omega\left(\varphi,\theta^2/a_1^2a_2^2\right)$ and thus the
compatibility condition (\ref{comp0}) becomes
\begin{equation}
B \, \Omega_{,\, \varphi} + C \, \Omega_{, \, \theta} =0\,\,\, ,
\label{comp1}
\end{equation}
which is independent of $A$. The constant of motion associated to
$X_{\perp},X_\varphi, X_{\theta}$ are the following:
\begin{eqnarray}
&&K_\perp=-M^2\, a_{3} \left(a_{1}\, \dot{a_{2}} - a_{2} \, \dot{a_{1}}\right) \,\,\, ,\\
&&K_\theta
= \frac{a_{L}}{a_{1} \,  a_{2}} \, \dot{\theta}\,\,\, , \\
&&K_\varphi = a_{1}\, a_2 \, a_{L}\, \dot{\varphi}\,\,\, .
\label{2.18}
\end{eqnarray}
in particular condition (\ref{comp1}) implies that $K_\perp$ is
always preserved for any $A\not=0 $. Moreover, in order to satisfy
Eq. (\ref{comp1}) we have the following possibilities only:
\\
\\ a) $\Omega=$ const. In this case we have three free parameters, namely $A$, $B$
and $C$ and thus three constants of motion $K_\perp$, $K_\varphi$
and $K_\theta$;
\\
\\
b) $B=0$, $\Omega=\Omega(\varphi)$, and two generic values for $A$
and $C$. In this case one has the two constants of motion
$K_\perp$ and $K_\theta$.
\\
\\
c) $C=0$ and $\Omega(\theta)$, and two generic values for $A$ and $B$.
Hence two constants of motion $K_\perp$ and $K_\varphi$.

The presence of constants of motion allows to reduce the dimension of the
configuration space and thus simplifies the dynamics of the system.

Let us consider as a relevant but simple model the case of a
cosmological constant $\Omega = V + W=const.$. In this case, the
dynamical system can be reduced to a two dimensional one, and
solved by using the equation of motion for the not cyclic
variables together with the constraint energy equation which fixes
the initial conditions. In particular one can choose a local set
of coordinate transformations such that
\begin{equation}
q_1\equiv \log\left(a_{1}\right),\,\,\,\,\,\,\,\, q_2\equiv
\log\left( a_{1} \, a_2\right),\,\,\,\,\,\,\,\, a\equiv
\left(a_{1} \, a_2 \, a_{L}\right)^{1/3}.
\end{equation}
In terms of these new variables the Lagrangian of Eq.(\ref{lan1}) becomes
\begin{equation}
{\cal L} = a^3\left\{M^2\left[\dot{q}_2^2 +
\dot{q}_1^2 - \dot{q}_1 \dot{q}_2 - 3 \frac{\dot{a}}{a}\dot{q}_2\right] \, +
\frac{\dot{\varphi}^2}{2} + \, e^{-2 q_2}\frac{\dot{\theta}
^{2}}{2} - \Omega\right\}\,\,\, ,
 \label{2.80}
\end{equation}
which is independent of $q_1$, $\theta$ and $\varphi$. As stated before, we
have three dimensionless constants of motion
\begin{eqnarray}
\widehat{K}_\perp = \frac{K_\perp}{M^3}\equiv \frac{1}{M^3} \, \frac{\partial{\cal L}}{\partial
\dot{q}_1} & = & \frac{1}{M} \, a^3 \, \left[ 2\dot{q}_1- \dot{q}_2 \right]\,\,\, ,\\
\widehat{K}_\varphi = \frac{K_\varphi}{M^2}\equiv \frac{1}{M^2} \, \frac{\partial{\cal L}}{\partial
\dot{\varphi}}&
= & \frac{1}{M^2} \, a^3\, \dot{\varphi} \,\,\, ,\\
\widehat{K}_{\theta}=\frac{K_{\theta}}{M^2} \equiv \frac{1}{M^2} \,\frac{\partial{\cal L}}{\partial
\dot{\theta}}& = & \frac{1}{M^2} \, a^3 \, e^{-2q_2}\, \dot{\theta}\,\,\, .
\end{eqnarray}
Hence we have
\begin{eqnarray}
\dot{q}_1&=& \frac{1}{2}\left[ \dot{q}_2 +
\frac{M \, \widehat{K}_\perp}{a^3}\right]\,\,\, , \label{3.61}
\\
\dot{\varphi}&=& M^2 \, \frac{\widehat{K}_\varphi}{a^3}\,\,\, , \label{3.62}
\\
\dot{\theta}&=& M^2 \, \frac{\widehat{K}_{\theta}}{a^3} \, e^{2q_2}\,\,\, .\label{3.6}
\end{eqnarray}
Therefore, we have reduced the problem to a 2-dimensional one, since once
$a(t), q_{2}(t)$ are found it is possible to get $q_{1}(t),
\varphi(t),
\theta(t)$ by direct integration of the previous equations. If we introduce the dimensionless
time $\tau \equiv M \, t$ and potential $\omega \equiv \Omega/M^4$, the
remaining set of equations for $a(\tau)$ and $q_2(\tau)$ to be solved now
reads
\begin{eqnarray}
&&q_2'' = - \frac 34 q_2'\,^2 + \omega + \frac{1}{2 \, a^6}
\left[\frac{1}{2} \, \widehat{K}_\perp^2 +
\widehat{K}_\varphi^2 + \widehat{K}_{\theta}^2 \,
e^{2q_2}\right]\,\,\, ,
 \label{q2dotdot}\\
&&\frac{a''}{a}=- 2 \left( \frac{a'}{a} \right)^2 + \frac 32
\left( \frac{a'}{a} \right) \, q_2' - \frac 38 q_2'\,^2 +
\frac{\omega}{2} + \frac{1}{4 \, a^6} \left[\frac{1}{2} \,
\widehat{K}_\perp^2 + \widehat{K}_\varphi^2 -
\frac{1}{3}\widehat{K}_{\theta}^2 \, e^{2q_2}\right],
\,\,\,\,\,\,\,\,\,\,\,\,\,\, \label{adotdot}
\end{eqnarray}
where the {\it prime}-index stands for $d/d\tau$.\\ A comment is
in turn. The quantity $\widehat{K}_\perp= a^3 | H_1 - H_2|/M$
where $H_1 \equiv \dot{a}_1/a_{1}$ and  $H_2 \equiv
\dot{a}_2/a_{2}$ is a measure of the initial spatial anisotropy in
the 1-2 plane. For the solutions with increasing $a=a(t)$, the
constancy of $\widehat{K}_\perp$ implies that $| H_1 - H_2|
\rightarrow 0$. Thus we have an expanding universe that becomes
asymptotically isotropic along $1, 2$-directions and in the
meanwhile, due to the constancy of $\widehat{K}_{\theta}$, the
$\theta$-field, contributing to anisotropy goes to zero. The
isotropization is complete since it involves $a_L$ as well. This
can be easily understood by observing that for increasing $a=a(t)$
the contribution of the square brackets to the r.h.s. of Eq.s
(\ref{q2dotdot}) and (\ref{adotdot}) asymptotically vanishes. In
this condition $q_2' \rightarrow 2 (a'/a)$ and thus $H_1 = H_2 =
H_L$ which means complete spatial isotropization.

\begin{figure}
\begin{center}
\epsfig{height=8truecm,width=10truecm,file=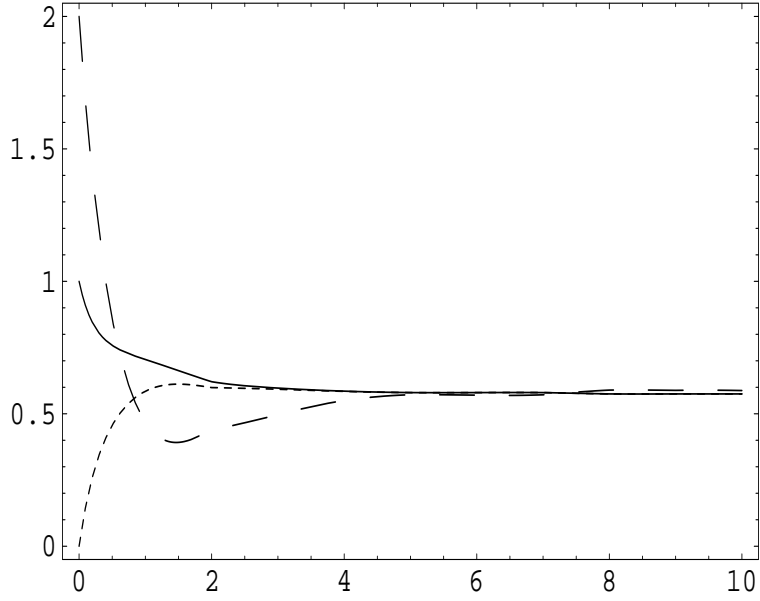}
\caption{The Hubble parameters along the direction $1$ and $2$,
namely $H_{1}$ (solid line),  $H_{2}$ (dashed line) are here plotted versus
$\tau$. It is also reported as a long-dashed line the behaviour of  $H_L$.
The evolution corresponds to the initial conditions
$a(0)=1,a'(0)=1,q_2(0)=1, q'_2(0)=1$.}
\label{fig1}
\end{center}
\end{figure}

Figure \ref{fig1} shows the  behaviour of the three directional
Hubble parameters $H_1(\tau)$, $H_2(\tau)$, and $H_L(\tau)$ versus
$\tau$, respectively. The evolution corresponds to the initial
conditions $a(0)=1,a'(0)=1,q_2(0)=1, q'_2(0)=1$ but the main
features shown for these particular choice of initial conditions
are quite general. In particular, as can be seen by looking at the
behaviour of $H_1$, $H_2$ and $H_L$, the presence of the
cosmological constant $\omega$ asymptotically yields to a complete
spatial isotropization and thus to a space-time symmetry
restoration.

\begin{figure}
\begin{center}
\epsfig{height=8truecm,width=10truecm,file=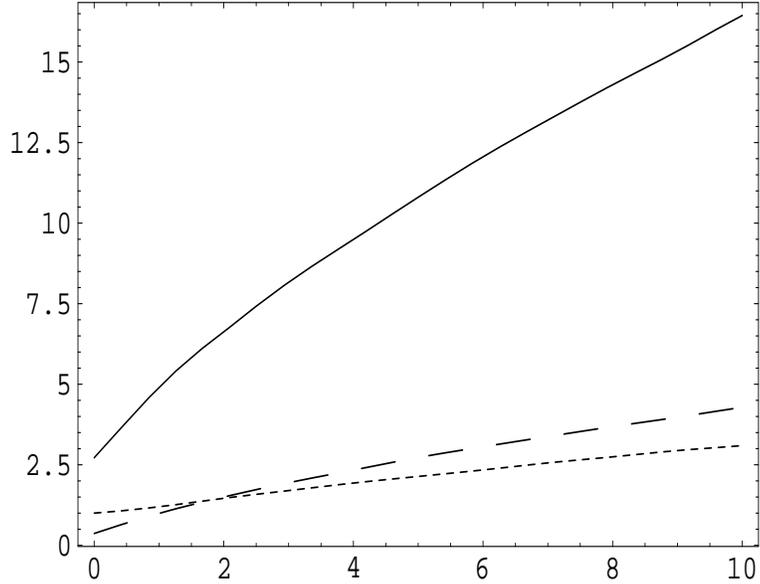}
\caption{The scale factors $a_1(\tau)$ (solid line) $a_2(\tau)$
(dashed line) and $a_3(\tau)$ (long-dashed line) are here plotted versus
$\tau$. The evolution corresponds to the initial conditions
$a(0)=1,a'(0)=1,q_2(0)=1, q'_2(0)=1$ .}
\label{fig2}
\end{center}
\end{figure}

\begin{figure}
\begin{center}
\epsfig{height=8truecm,width=10truecm,file=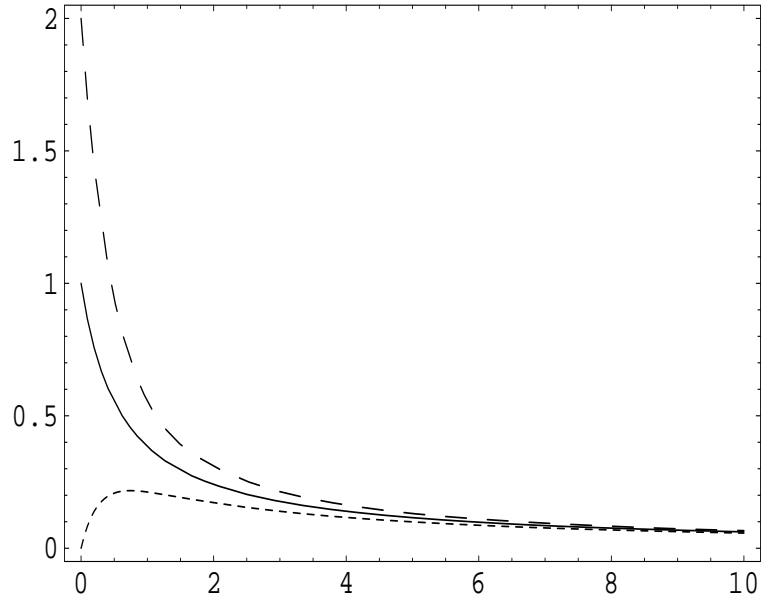}
\caption{The Hubble parameters along the direction $1$ and $2$,
namely $H_{1}$ (solid line),  $H_{2}$ (dashed line) are here plotted versus
$\tau$. It is also reported as a long-dashed line the behaviour of  $H_L$.
The evolution corresponds to the initial conditions
$a(0)=1,a'(0)=1,q_2(0)=1, q'_2(0)=1$.}
\label{fig3}
\end{center}
\end{figure}

As a second relevant case we can assume for the potential $\Omega$ in the
lagrangian density (\ref{lan1}) the following form
\begin{equation}
\Omega\left(\varphi,\theta_{\mu\nu}\theta^{\mu\nu}\right)= \xi \,
\varphi^2 \, \theta_{\mu\nu}\theta^{\mu\nu} = \xi \, \varphi^2
\frac{\theta^2}{a_1^2a_2^2}\,\,\, .\label{pot2}
\end{equation}
In this case $K_\perp$ is the only constant of motion.\\ As in the
previous case, the Lagrangian of Eq.(\ref{lan1}) becomes
\begin{equation}
{\cal L} = a^3\left\{M^2\left[\dot{q}_2^2 +
\dot{q}_1^2 - \dot{q}_1 \dot{q}_2 - 3 \frac{\dot{a}}{a}\dot{q}_2\right] \, +
\frac{\dot{\varphi}^2}{2} + \, e^{-2q_2}\frac{\dot{\theta}
^{2}}{2} - \xi e^{-2q_2} \, \varphi^2 \, \theta^2\right\}\,\,\, ,
 \label{lann2}
\end{equation}
Therefore, we have reduced the problem to a 4-dimensional one, since once
$a(t)$, $q_{2}(t)$, $\varphi(t)$, $\theta(t)$ are found it is possible to
get $q_{1}(t)$ by direct integration of eq. (\ref{3.61}). As in the
previous case, we introduce the dimensionless time $\tau \equiv M \, t$ and
$\tilde{\varphi}(\tau)=M\varphi,
\tilde{\theta}(\tau)=\theta$, the remaining set of equations for
$a(\tau), q_2(\tau),\tilde{\varphi}(\tau), \tilde{\theta}(\tau)$
to be solved now reads:
\begin{eqnarray}
\varphi'' &=& -2\xi e^{-2q_2} \, \tilde{\varphi} \,
\tilde{\theta}^2 - 3\tilde{\varphi}'\frac{a'}{a}\,\,\, ,
\\
\tilde{\theta}'' &=& -2\xi \, \tilde{\varphi}^2 \,
\tilde{\theta} - 3\tilde{\theta}'\frac{a'}{a} +
2q_2'\tilde{\theta}'\,\,\, ,
\\
q_2'' &=& - \frac 34 q_2'\,^2 - \frac{1}{2} \tilde{\varphi}'^{2}  + \left[
\frac{1}{4 \, a^6}
\widehat{K}_\perp^2 -\frac{1}{2} \tilde{\theta}'^{2} \, e^{-2q_2} + \xi
e^{-2q_2} \, \tilde{\varphi}^2 \, \tilde{\theta}^2 \right]
\,\,\, ,
 \label{q2dotdot2}\\
\frac{a''}{a}&=&- 2 \left( \frac{a'}{a} \right)^2 + \frac 32
\left( \frac{a'}{a} \right) \, q_2' - \frac 38 q_2'\,^2
-\frac{1}{4}\tilde{\varphi}'^{2}
 \nonumber\\ &+& \left[\frac{1}{8 \, a^6}  \, \widehat{K}_\perp^2
+\frac{1}{12}\tilde{\theta}'^{2}\, e^{-2q_2}
-\frac{1}{6} \xi e^{-2q_2} \, \tilde{\varphi}^2 \, \tilde{\theta}^2\right]
\,\,\, .\label{adotdot2}
\end{eqnarray}
Figures \ref{fig2} and \ref{fig3} show the  behaviour of $a_1(\tau)$,
 $a_2(\tau)$, $a_L(\tau)$ and $H_1(\tau)$, $H_2(\tau)$, $H_L(\tau)$ versus $\tau$.
The evolution corresponds to $\xi=0.5$ and to the initial
conditions $a(0)=1$, $a'(0)=1$, $q_2(0)=1$, $q'_2(0)=1$ but the
main features shown are quite general. Even in this case, the
contribution of the square brackets to the r.h.s. of Eq.s
(\ref{q2dotdot2}) and (\ref{adotdot2}) asymptotically vanishes and
thus $q_2' \rightarrow 2 (a'/a)$ and thus $H_1 = H_2= H_L$ which
again yields to complete spatial isotropization.

To conclude we have studied the space-time symmetry restoration in
a cosmological model where a minimally coupled scalar field is
interacting with a K-R field. The analysis has been performed in a
Bianchi I universe and restricted to the simplest cases of a
cosmological constant (not interacting fields) and in presence of
a pure quadratic and renormalizable interaction term. In both
cases, even if the two systems are characterized by a different
number of constants of motion and thus by a different level of
dynamical symmetry, an asymptotically complete spatial symmetry
restoration is obtained. This is due to the universe expansion
which dilutes the contribution of the interaction terms
responsible for a possible spatial anisotropy.

The authors would like to thank G. Marmo for useful discussion.


\begin{thebibliography}{99}
\bibitem{Bennett:2003bz}
C.~L.~Bennett {\it et al.}, Astrophys.\ J.\ Suppl.\  {\bf 148}, 1
(2003).
\bibitem{Lizzi:2002ib}
F.~Lizzi, G.~Mangano, G.~Miele and M.~Peloso, JHEP {\bf 0206}, 049
(2002).
\bibitem{DiGrezia:2003ug} E.~Di Grezia, G.~Esposito, A.~Funel, G.~Mangano and G.~Miele,
Phys.\ Rev.\ D {\bf 68}, 105012 (2003).
\bibitem{deRitis:ba}
R.~de Ritis, G.~Marmo, G.~Platania, C.~Rubano, P.~Scudellaro and
C.~Stornaiolo, Phys.\ Rev.\ D {\bf 42}, 1091 (1990).
\bibitem{deRitis:tp}
R.~de Ritis, G.~Marmo, G.~Platania, C.~Rubano, P.~Scudellaro and
C.~Stornaiolo, Phys.\ Lett.\ A {\bf 149}, 79 (1990).
\bibitem{Capozziello:bi}
S.~Capozziello, R.~De Ritis, C.~Rubano and P.~Scudellaro, Riv.\
Nuovo Cim.\  {\bf 19N4} (1996) 1.
\bibitem{dem} M.~Demianski et al. Phys.\ Rev. D {\bf 46} 1391
(1992).
\end{thebibliography}
\end{document}